\documentclass[aps,prd,twocolumn,superscriptaddress]{revtex4-1}

\usepackage[english]{babel}
\usepackage{amsmath, amsfonts, amsthm, amsbsy, amsopn, amsxtra, amscd, amstext}
\usepackage{longtable}
\usepackage{bbm}
\usepackage{enumitem}
\usepackage{graphicx}
\usepackage{xcolor}
\usepackage{wrapfig}
\usepackage{soul}
\usepackage{physics}
\usepackage{amssymb}
\usepackage[hidelinks]{hyperref}
\def\apj{Astrophysical Journal}                %
\def\mnras{MNRAS}            %
\def\prd{Phys.~Rev.~D}       %
\def\prl{Phys.~Rev.~Lett}    %

\newcommand\SkipFisher[1]{}

\newcommand{\IMRPD}{\textsc{IMRPhenomD}}
\newcommand\RIFT{RIFT}

\newcommand\mc{{\cal M}_c}
\newcommand\HideMe[1]{}
\newcommand\unit[1]{{\rm #1}}

\newcommand{\AffiliationCCRG}{%
  Center for Computational Relativity and Gravitation, %
  Rochester Institute of Technology, %
  Rochester, New York 14623, USA %
}
\newcommand{\AffiliationNASA}{%
NASA Postdoctoral Program, Astrophysics Science Division, NASA Goddard Space Flight Center, Greenbelt, MD 20771, USA}

\begin{document}
\title{
 Low-latency parameter inference enabled by a Gaussian likelihood approximation for RIFT
}

\author{A. B. Yelikar}
\affiliation{\AffiliationCCRG}
\author{V. Delfavero}
\affiliation{\AffiliationCCRG}
\affiliation{\AffiliationNASA}
\author{R. O'Shaughnessy}
\affiliation{\AffiliationCCRG}


\date{\today}
\begin{abstract}
Rapid identification, characterization, and localization of gravitational waves from binary compact object mergers
can enable well-informed follow-on multimessenger observations.   
In this work, we investigate a small modification to the RIFT parameter inference pipeline to enable extremely
low-latency inference, tested here for nonprecessing sources.

\end{abstract}

\maketitle

\section{Introduction}
\label{sec:intro}

The discovery and sky localization of gravitational waves (GW) from  GW170817 enabled a broad-based multimessenger followup
campaign with transformative implications for our understanding of binary mergers.  
As ground-based GW detector networks will only increase in sensitivity,  the growing number of multimessenger
candidates during the next observing run (O4) and beyond increases the urgency  for rapid, ubiquitous, and reliable parameter inference of
compact binary mergers.  
These rapid inferences provide not only telescope pointing information, with sky localization and distance \cite{LIGO-O2-EMFollowup}, but also
information to characterize the nature of the sources (masses, spins), which can be vital in determining if an object
could plausibly produce an EM counterpart  \cite{LIGO-2013-WhitePaper-CoordinatedEMObserving}.

Many methods for rapid parameter inference exist, often adopting different tradeoffs between speed and accuracy.
BayesStar\cite{Singer_2016}, the most widely used in low-latency alerts, originally focused on sky location (and distance estimates),
fixing its estimate of source parameters to search results, providing results within seconds.
Other longer-latency methods attempt to simultaneously localize the source and measure its parameters. 
These methods often achieve short turnaround time by aggressive approximations, including 
fast waveform models, accelerated likelihoods, and particularly simplified physics (e.g., no precession or higher-order
modes).
The last few years have seen an explosion of improved tools to enable low-latency inference.  Some groups have
developed  specialized
infrastructure for existing Bayesian inference engines, such as fast waveforms
\cite{2022PhRvD.106j4029T,2022arXiv221015684T,2021PhRvD.104h4058G}, modified likelihoods 
(e.g., reduced order quadrature \cite{gwastro-mergers-PE-ReducedOrder-2013,gwastro-pe-ROM-IMRPv2-2016,Morisaki:2020oqk}, 
multibanded likelihoods \cite{2017CQGra..34k5006V,Morisaki:2021ngj}, and relative binning
\cite{2018arXiv180608792Z,2021PhRvD.104j4054C}), or 
alternative sampling \cite{2021PhRvD.103j3006W,gwastro-mergers-nessaiv2-2022}.
Other approaches are inspired by RIFT's architecture
\cite{gw-astro-mergers-VTiwariPE,gw-astro-mergers-cogwheel2022}, while the most aggressive options use machine learning
methods (albeit so far only for massive black holes) \cite{2021PhRvL.127x1103D,gw-astro-mergers-dingo-part2,2022NatPh..18..112G}.
One extremely promising approach closely related to RIFT uses an iteratively \emph{refined} grid of intrinsic points to
identify the most-appropriate intrinsic parameters
\cite{2022arXiv220105263R}.






\nocite{2021MNRAS.507.2037A} 


\nocite{2022arXiv220105263R,gwastro-RIFT-Update}

In this work, we investigate a small  modification to the RIFT parameter inference pipeline to enable extremely
low-latency inference, tested here for nonprecessing sources.
While the RIFT parameter inference pipeline can
produce fairly rapid inference, based on fast interpolated generic likelihoods, the computational cost needed to
guarantee a fully-explored and very accurate likelihood (and posterior) over all pertinent physical can be prohibitive.
In this work,  we  limit the model dimension and adopt a modest but pertinent approximation: nearly-Gaussian
(intrinsic) likelihoods.  For simplicity and given observationally pertinent considerations -- most neutron stars are
in binaries whose components are consistent with small or zero spin  -- in this work we only investigate parameter
inference of nonprecessing binaries, allowing us to substantially reduce the pertinent intrinsic dimension.

This paper is organized as follows.  In Section  \ref{sec:Methods} we outline our parameter inference engine (RIFT),
Gaussian likelihood approximations within it, and the synthetic population of events we'll use to test our conclusions.
In Section \ref{sec:results} we demonstrate by example our approach reliably reproduces the key intrinsic and extrinsic
parameters of our synthetic population.

This investigation is important not because we provide yet another high-performing low-latency inference strategy, but rather that this
approach builds on (and interfaces naturally with) the well-tested RIFT pipeline, automatically inheriting its
capabilities.  Thus, this technique can use models higher-order modes or time-domain models, on the one hand, and efficiently inform parallel
(and more robust) \emph{generic}-likelihood analyses, on the other.  

\section{Methods}
\label{sec:Methods}

\subsection{RIFT review}
\label{subsec:RIFT}

A coalescing compact binary in a quasicircular orbit can be completely characterized by its intrinsic
and extrinsic parameters.  By intrinsic parameters, we refer to the binary's  masses $m_i$, spins, and any quantities
characterizing matter in the system.  For simplicity and reduced computational overhead, in this work we assume all
compact object spins are aligned with the orbital angular momentum. 
By extrinsic parameters, we refer to the seven numbers needed to characterize its spacetime location and orientation.  
We will express masses in solar mass units and
 dimensionless nonprecessing spins in terms of cartesian components aligned with the orbital angular momentum
 $\chi_{i,z}$.   We will use $\lambda,\theta$ to
refer to intrinsic and extrinsic parameters, respectively.

RIFT \cite{gwastro-PENR-RIFT}
consists of a two-stage iterative process to interpret gravitational wave data $d$ via comparison to
predicted gravitational wave signals $h(\lambda, \theta)$.   In one stage, for each  $\lambda_\alpha$ from some proposed
``grid'' $\alpha=1,2,\ldots N$ of candidate parameters, RIFT computes a marginal likelihood 
\begin{equation}
 {\cal L}_{\rm marg}\equiv\int  {\cal L}(\boldsymbol{\lambda} ,\theta )p(\theta )d\theta
\end{equation}
from the likelihood ${\cal L}(\bm{\lambda} ,\theta ) $ of the gravitational wave signal in the multi-detector network,
accounting for detector response; see the RIFT paper for a more detailed specification.  
In the second stage,  RIFT performs two tasks.  First, it generates an approximation to ${\cal L}(\lambda)$ based on its
accumulated archived knowledge of marginal likelihood evaluations 
$(\lambda_\alpha,{\cal L}_\alpha)$. This approximation can be generated by gaussian processes, random forests, or other
suitable approximation techniques. Second, using this approximation, it generates the (detector-frame) posterior distribution
\begin{equation}
\label{eq:post}
p_{\rm post}(\boldsymbol{\lambda})=\frac{{\cal L}_{\rm marg}(\boldsymbol{\lambda} )p(\boldsymbol{\lambda})}{\int d\boldsymbol{\lambda} {\cal L}_{\rm marg}(\boldsymbol{\lambda} ) p(\bm{\lambda} )}.
\end{equation}
where prior $p(\bm{\lambda})$ is the prior on intrinsic parameters like mass and spin. The posterior is produced by
performing a Monte Carlo integral:  the evaluation points and weights in that integral are weighted posterior samples,
which are fairly resampled to generate conventional independent, identically-distributed ``posterior samples.''
For further details on RIFT's technical underpinnings and performance, see
\cite{gwastro-PENR-RIFT,gwastro-PENR-RIFT-GPU,gwastro-mergers-nr-LangePhD}.

\subsection{Gaussian likelihoods for RIFT}

Normally, RIFT builds the approximation $\hat{{\cal L}}(\lambda)$ using a very flexible nonparametric interpolator, such as a
Gaussian process or random forest.  With a Gaussian approximation, however, we instead require $\ln\hat{{\cal L}}$ is
a quadratic polynomial, which can always be recast as
\begin{align}
\label{eq:quad}
\ln \hat{\cal L}_{cov} = \ln {\cal L}_{maxfit} -  \frac{1}{2} (\lambda - {\lambda}_*)_p (\lambda - \lambda_*)_q \Gamma_{pq}
\end{align}
where $\ln {\cal L}_{\rm maxfit},\lambda_*,\gamma$ are all identified by the quadratic fit.
In suitable coordinates, a quadratic fit of this form can be an extremely reliable representation of the near-peak
likelihood, even when including most of the pertinent intrinsic and extrinsic dimensions
\cite{nal-chieff-paper,nal-methods-paper,DelfaveroMastersThesis}. 
For simplicity, however, in this work we will only employ quadratic fits using three intrinsic coordinates: the chirp mass
$\mc=(m_1m_2)^{3/5}/(m_1+m_2)^{1/5}$, the symmetric mass ratio $\eta=m_1m_2/(m_1+m_2)^2$, and the effective inspiral
spin $\chi_{\rm eff} = (m_1 \chi_{1} + m_2 \chi_2)\cdot \hat{\bf L}/(m_1+m_2)$, where $m_i$ are the component masses,
$\chi_i$ the component spins, and $\hat{ \bf L}$ is the (Newtonian) orbital angular momentum direction.

To avoid distorting our near-peak fit with off-peak information, the appropriate quadratic form is identified using only
likelihood data within 20\% of the peak likelihood.  We have developed a few strategies to identify the
specific parameters of the quadratic from our training data, to minimize numerical issues associated with the very
narrow posterior (and thus potentially ill-conditioned matrices).  For this work, however, we simply perform a
least-squares fit to   $\ln {\cal L}$ versus $\bm\lambda$, after scaling the latter to zero mean and unit variance to
stabilize our numerical methods.

From the Gaussian likelihood, we then construct posterior samples by first drawing fair samples $\lambda'_\beta$ from
the  \emph{likelihood} (i.e., $\lambda'_\beta$ are normal with mean $\lambda_*$ and covariance $\Gamma^{-1}$);
rejecting draws outside pertinent boundaries; 
weighting those samples by the prior $p(\lambda'_\beta)$; and then resampling to produce a set of fair posterior draws.
For example and for simplicity ignoring changes in normalization associated with finite boundaries, with this procedure we estimate the evidence integral as follows:
\begin{align}
\int {\cal L}(\lambda) p(\lambda) d\lambda = {\cal L}_{\rm max} \frac{(2\pi)^{d/2}}{|\Gamma|^{d/2}} \frac{1}{N} \sum_k p(\lambda_k)
\end{align}


\subsection{Low-latency Gaussian-RIFT strategy}
Rather than employ the full complex infrastructure used for a typical RIFT pipeline, we instead propose a simple
two-step Gaussian-RIFT (G-RIFT) low-latency strategy for low-mass binaries.  First, propose a (random) initial grid of intrinsic points,
following usual RIFT conventions \cite{gwastro-RIFT-Update}.  Approximate the likelihood by a gaussian, generate
candidate points, re-evaluate with the gaussian, regenerate candidate points, and then use those final candidate points
as the overall intrinsic posterior, generating extrinsic posterior samples as usual for RIFT  \cite{gwastro-RIFT-Update}.

\subsection{Synthetic population of events}
\label{sec:sub:pop}

We explore this technique over a limited fiducial population.
Specifically, we consider a universe of synthetic signals for 
3-detector networks, with masses
drawn uniformly in $m_i$ in the region bounded by $\mc/M_\odot \in [1.2,1.4 ]$ and $\eta \in [0.2, 0.25]$ and with
extrinsic parameters drawn uniformly in sky position and isotropically in Euler angles, with source luminosity
distances  drawn proportional to $d_L^2$  between
$30\unit{Mpc}$ and $200\unit{Mpc}$. 
These bounds are expressed in terms of $\mc=(m_1m_2)^{3/5}/(m_1+m_2)^{1/5}$ and $\eta=m_1m_2/(m_1+m_2)^2$, 
and encompass the detector-frame parameters of neutron stars seen in GWTC-1 \cite{GWTC-1}, 
GWTC-2 \cite{GWTC-2}, GWTC-2.1 \cite{GWTC-2p1} and GWTC-3 \cite{GWTC-3}. 
Unless otherwise noted, all our sources have no tides, with both spins nonprecessing and with $z$ components uniformly distributed in $\chi_{i,z}\in[-0.05,0.05]$.
%
For complete reproducibility, we use \texttt{\IMRPD}, starting the signal evolution and likelihood integration 
at $30\unit{Hz}$, performing all analysis with $4096\unit{Hz}$ time series in Gaussian noise with 
known advanced LIGO design PSDs \cite{LIGO-aLIGODesign-Sensitivity-Updated}. 

One way to assess the performance of parameter inference is a probability-probability plot (usually denoted PP plot) \cite{mm-stats-PP}.
  Using \RIFT{} on each source $k$, with true parameters $\mathbf{\lambda}_k$, we estimate
the fraction of the posterior distributions which is below the true source value $\lambda_{k,\alpha}$   [$\hat{P}_{k,\alpha}(<\lambda_{k,\alpha})$] 
for each intrinsic parameter $\alpha$, again assuming all sources
are aligned and slowly spinning.  After reindexing the sources so $\hat{P}_{k,\alpha}(\lambda_{k,\alpha})$ increases with $k$ for some
fixed $\alpha$,  we make a
plot of $k/N$ versus $\hat{P}_k(\lambda_{k,\alpha})$ for all binary parameters.


\section{Results}
\label{sec:results}

\subsection{Assessing reliability of inferences}

\begin{figure}
\includegraphics[width=\columnwidth]{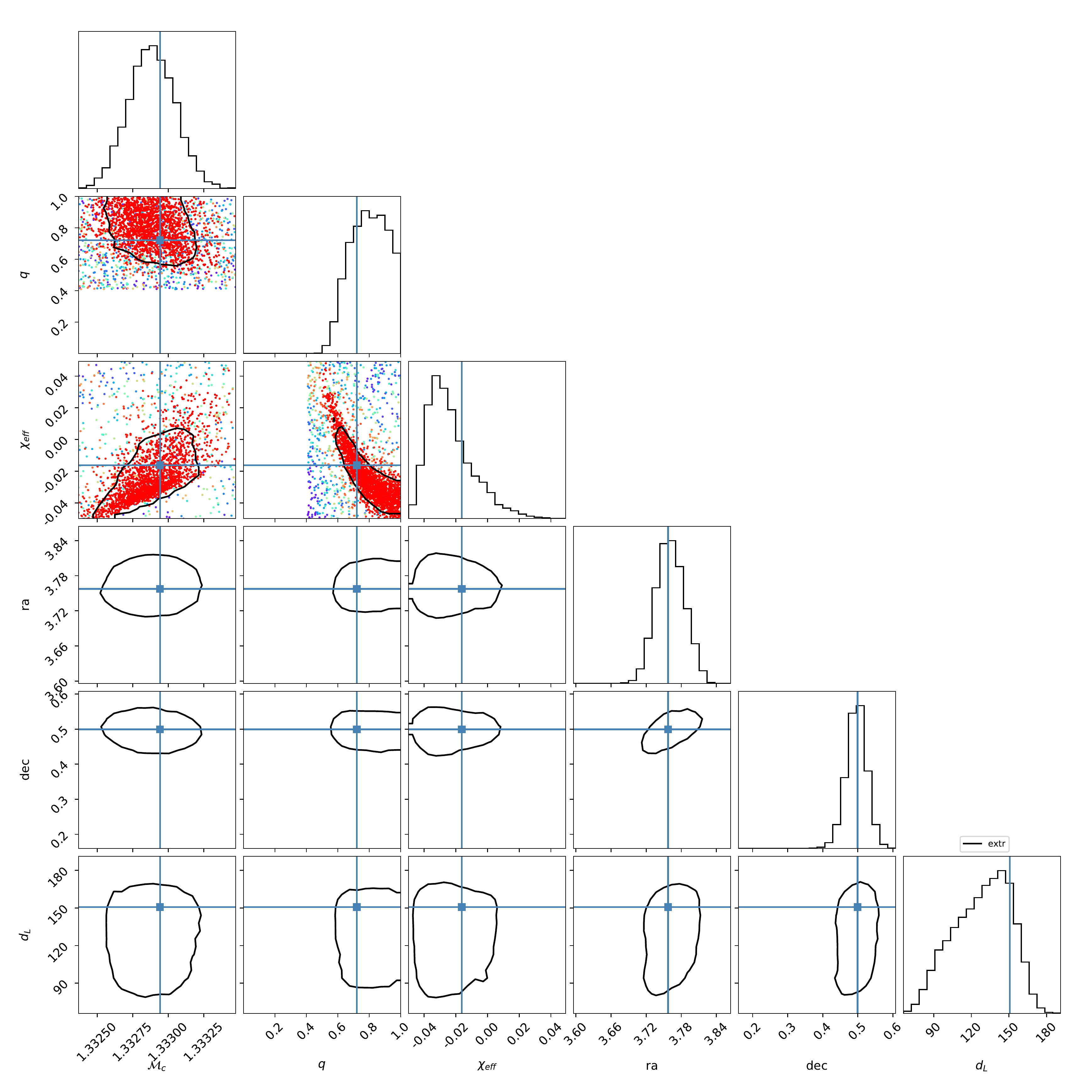}
\caption{\label{fig:fiducial_alignedspin} Fiducial event with aligned spin, from PP set. Colored points 
denote marginalized likelihood over the extrinsic parameters (red indicating a 
higher likelihood).}
\end{figure}

Figure \ref{fig:fiducial_alignedspin} shows an example of our two-stage process, applied to a synthetic binary neutron star
in Gaussian noise. Each point corresponds to a candidate set
of intrinsic parameters, where the color scale shows the inferred
(marginal) likelihood. The contours (and, on the diagonal, one-dimensional marginal distributions) represent the
posterior distribution.

\begin{figure}
\includegraphics[width=\columnwidth]{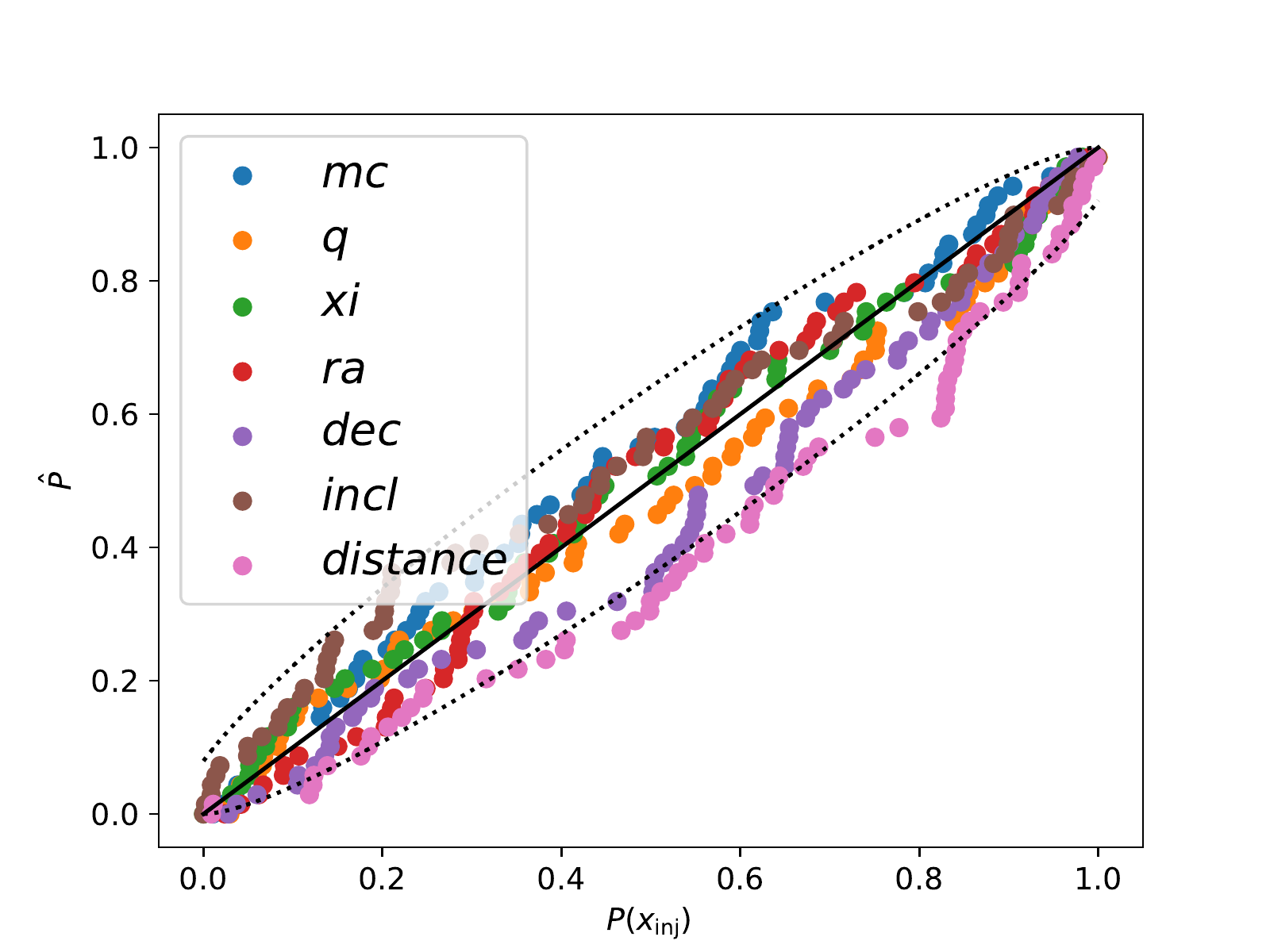}
\caption{\label{fig:pp_alignedspin} Probability-probability plot for spinning, nonprecessing binary neutron stars,
  including key intrinsic parameters ($\mc,q,\chi_{\rm eff}$) and key extrinsic properties (sky location and distance).
 The dashed line indicates the 90\% credible interval expected for a cumulative distribution drawn from the population
 of uniformly-distributed samples.}
\end{figure}

Going beyond anecdote, Figure \ref{fig:pp_alignedspin} shows a PP plot test for our two-stage process.  We see both the
key intrinsic and extrinsic parameters are well-recovered by this approach.

\subsection{Assessing inference timescales}

Even though we have not optimized any settings for this approach (e.g., the number of likelihood evaluations), this
method performs inference extremely rapidly.  As an example, the nonprecessing $(2,\pm 2)$-mode model only requires
2 second per likelihood evaluation, and 10 seconds to generate the posterior.  As a result,
the two iterations of  $2\times10^3$ likelihood evaluations each only require $10$ seconds each on 400 GPU cores,
implying the inference \emph{could} be performed in as little as 50 seconds, barring resource contention (10+10+10+10+10,
allowing for a final 10-second evaluation of the extrinsic parameters). Our runtime estimate can easily be reduced by
reducing the (substantial) redundancy in our followup grids, and by otherwise reducing overhead.



\section{Conclusions}
Working within the RIFT GW source parameter inference pipeline, we have demonstrated a technique to rapidly infer
the properties of candidate low-mass binary mergers.   Our method complements the even faster adaptive-mesh-refinement
technique introduced in \cite{2022arXiv220105263R}, which performs much  fewer likelihood evaluations and employs a
streamlined pipeline.  Like that AMR method, this approach works within the RIFT ecosystem -- we only swapped  one
executable -- and therefore benefits from all ongoing developments to improve its speed and performance
\cite{gwastro-RIFT-Update}.  While the analysis presented here only used a fast, quadrupole-only approximation, this
approach will work equally well for any approximation, including models with higher order modes.  Additionally, given
how well Gaussian approximations seem to capture the properties of even precessing sources \cite{nal-methods-paper,DelfaveroMastersThesis}, we
anticipate this approach may also enable fast inference for precessing sources as well. 
Finally, in addition to potentially providing low-latency inference directly, this method provides a 
potential bootstrap for modestly longer-duration analyses with more comprehensive physics and systematics.

While we test our Gaussian approach on a wide range of synthetic sources, including marginally significant events, we
anticipate our Gaussian method will behave most poorly for the lowest-amplitude event candidates, whose likelihoods  are
least-well-approximated by a Gaussian. That said, our method (and the AMR technique \cite{2022arXiv220105263R}) could be
run in parallel with a bootstrapped more robust RIFT architecture to provide guaranteed modest-latency inference while
enabling fast inference in the frequent cases where a Gaussian approximation suffices.

\begin{acknowledgments}
ROS, VD, and AY have been supported by NSF-PHY 2012057.
    ROS is also supported via NSF PHY-1912632 and AST-1909534.
VD is supported by an appointment to the NASA Postdoctoral Program at the NASA Goddard Space Flight Center administered by Oak Ridge Associated Universities under contract NPP-GSFC-NOV21-0031
This material is based upon work supported by NSF’s LIGO Laboratory 
    which is a major facility fully funded by the
    National Science Foundation.
This research has made use of data,
    software and/or web tools obtained from the Gravitational Wave 
    Open Science Center (https://www.gw-openscience.org/ ),
    a service of LIGO Laboratory,
    the LIGO Scientific Collaboration and the Virgo Collaboration.
LIGO Laboratory and Advanced LIGO are funded by the 
    United States National Science Foundation (NSF) as
    well as the Science and Technology Facilities Council (STFC) 
    of the United Kingdom,
    the Max-Planck-Society (MPS), 
    and the State of Niedersachsen/Germany 
    for support of the construction of Advanced LIGO 
    and construction and operation of the GEO600 detector.
Additional support for Advanced LIGO was provided by the 
    Australian Research Council.
Virgo is funded through the European Gravitational Observatory (EGO),
    by the French Centre National de Recherche Scientifique (CNRS),
    the Italian Istituto Nazionale di Fisica Nucleare (INFN),
    and the Dutch Nikhef,
    with contributions by institutions from Belgium, Germany, Greece, Hungary,
    Ireland, Japan, Monaco, Poland, Portugal, Spain.
The authors are grateful for computational resources provided by the 
    LIGO Laboratory and supported by National Science Foundation Grants
    PHY-0757058, PHY-0823459 and PHY-1626190 and IUCAA LDG cluster Sarathi.
\end{acknowledgments}

\footnotesize\bibliography{../Bibliography,%
gw-astronomy-mergers,%
gw-astronomy-mergers-approximations,%
LIGO-publications}



\end{document}